\documentclass[smallextended]{svjour3}
\usepackage{fix-cm}
\usepackage{bbm}
\usepackage{amsmath}
\usepackage{amssymb}
\usepackage{url}
\usepackage{longtable}
\usepackage{appendix}
\usepackage[vcentermath]{youngtab}
\def\IC{{\mathbb C}}
\def\IR{{\mathbb R}}

\def\cE{{\mathcal E}}

\def\cA{{\mathcal A}}

\def\0{{\mathbf 0}}

\def\({\left(}
\def\){\right)}
\def\[{\left[}
\def\]{\right]}

\def\int{{\bf Int}}

\def\>{{\rangle}}
\def\<{{\langle}}
\def\lf{{\lfloor}}
\def\rf{{\rfloor}}
\newcounter{countalg}
{\refstepcounter{countalg}%
\begin{minipage}{\textwidth}\baselineskip=1.5mm %
\setlength\unitlength{1truecm}%
\begin{picture}(15,0)(0,0)\put(0,0){\line(1,0){11.882}}\end{picture}\vspace{2mm}\\%
{\bf Algorithm \thecountalg}~#1\\\begin{quote}\vspace{-1mm}~}%
{\end{quote}%
\setlength\unitlength{1truecm}%
\begin{picture}(15,0)(0,0)\put(0,0){\line(1,0){11.882}}\end{picture}%
\end{minipage}}
{\begin{minipage}{\textwidth}%
\setlength\unitlength{1truecm}%
\begin{picture}(15,0)(0,0)\put(0,0){\line(1,0){11.882}}\end{picture}%
\baselineskip=1.5mm\begin{quote}}%
{\end{quote}%
\setlength\unitlength{1truecm}%
\begin{picture}(15,0)(0,0)\put(0,0){\line(1,0){11.882}}\end{picture}%
\end{minipage}}

\journalname{Quantum Information Processing}
\begin{document}
\title{Maximal noiseless code rates for
collective rotation channels on qudits}
\titlerunning{Maximal noiseless code rates for
collective rotation channels on qudits}
\author{Chi-Kwong Li \and Mikio Nakahara \and Yiu-Tung Poon \and
Nung-Sing Sze}
\institute{C.-K. Li \at
Department of Mathematics, College of William \& Mary,
Williamsburg, VA 23187-8795, USA. 
\and
M. Nakahara \at
Research Center for Quantum Computing,
Graduate School of Science and Engineering,
and Department of Physics,
Kinki University, 3-4-1 Kowakae, Higashi-Osaka, 577-8502, Japan.
\\
Department of Mathematics, and Department of Physics, Shanghai University,
200444 Shanghai, 
People's Republic of China.
\and
Y.-T. Poon \at
Department of Mathematics, Iowa State University,
Ames, IA 50011, USA.
\and
N.-S. Sze \at
Department of Applied Mathematics, The Hong Kong Polytechnic University,
Hung Hom, Hong Kong.
}

\date{Received: date / Accepted: date}

\maketitle

\begin{abstract}
We study noiseless subsystems on collective rotation channels of qudits, i.e.,
quantum channels with operators in the set
$\cE(d,n)  = \{ U^{\otimes n}:
U \in {\mathrm{SU}}(d)\}.$
This is done by analyzing
the decomposition of the algebra $\cA(d,n)$ generated by $\cE(d,n)$.
We summarize the results for the channels on qubits ($d=2$), and
obtain the maximum dimension of the noiseless subsystem that
can be used as the quantum error correction code for the channel.
Then we extend our results to general $d$. In particular, it is shown
that the code rate, i.e., the number of protected qudits
over the number of physical qudits, always approaches 1 for
a suitable noiseless subsystem.  Moreover, one can determine the maximum
dimension of the noiseless subsystem by solving a non-trivial
discrete optimization
problem. The maximum dimension of the noiseless subsystem for 
$d = 3$ (qutrits) is explicitly determined by
a combination of mathematical analysis and the symbolic software
Mathematica.
\end{abstract}



\section{Introduction}\label{secInt}

A quantum system is vulnerable to external noise. In quantum information
processing and quantum computation, the system must be protected
from the environmental noise one way or another
to protect quantum information stored in qubits.

Suppose $n$ photons are sent through an optical fiber with fixed imperfections
that cause polarization rotation of each photon
\cite{Holbrook,BRS}.
Then all photons suffer from the same error operator $U$,
which results in the collective rotation operator $U^{\otimes n}$.
Such an error operator is symmetric under an arbitrary
element of the symmetry group $S_n$ and one can take
advantage of this fact to find a subspace/subsystem immune to the
collective rotation operators. If quantum information is encoded in
this subspace/subsystem, it is protected from the noise.

The same situation appears when Alice sends $n$-qubits to
Bob without a common reference frame \cite{BRS0,BRS}. If Alice employs a
basis $\{ |0 \>, |1 \>\}$ while Bob employs a basis $\{
|0' \>, |1' \>\}$ for each qubit, these two bases are related by a unitary
rotation $U$. 
Then the frames of $n$-qubit system are different by $U^{\otimes n}$
and this ``mismatching'' of the frames are regarded as a collective rotation.
It can be shown that the same scheme also allows classical information
transmission without a shared reference frame \cite{BRS0,BRS}.

For implicitly it is assumed in the above two examples that the number of
qubits does not change during transmission. It can be shown that
quantum information can be protected even under 
quantum erasure noise, in which a photon is lost during transmission,
if the information is encoded in a suitable subspace \cite{MB,MZLWJ,CGL}.  
By the same token, one may distribute quantum information among
$n$ parties in such a way that any $k (<n)$ parties can reconstruct the
quantum information precisely but any $k-1$ parties cannot get any
information concerning the quantum information (quantum secret sharing).

Suppose $n$-qubit quantum states
$\rho$ are represented as $N\times N$ density matrices with $N = 2^n$,
and a quantum channel is realized as a completely positive linear map
$\Phi$ with an operator sum representation (OSR)
\begin{equation} \label{channel}
\Phi(\rho) = \sum_{i=1}^k E_i \rho E_i^{\dag}
\end{equation}
for the error operators $E_1, \dots, E_k$; see \cite{NO,NC}.
Then the error operators of our channel can be expressed as tensor products 
of error operators of the form $W^{\otimes n}$, $W \in {\rm SU(2)}$.

Decoherence free subspace and noiseless subsystem are standard
methods to avoid collective errors; see \cite{Zan1,Zan2,Zan3,YBG,Kempe,Kribs,Lidar}.
It is not hard to explain the scheme using the OSR
of the quantum channel (\ref{channel}) as follows. Suppose the
finite-dimensional $C^*$-algebra $\cA$
generated by the error operators admits the unique decomposition up to unitary
equivalence (similarity) as
$$\bigoplus_{j} (I_{s_j} \otimes M_{n_j}) \qquad \hbox{ with }
\ \sum_j s_j n_j = N.$$
In other words, there is a unitary matrix $Q \in M_N$
such that for every error operator $E_i$ in (\ref{channel}),
$$Q^{\dag}E_i Q = \bigoplus_{j} (I_{s_j} \otimes B_j^{(i)}) \qquad
\hbox{ with } \ B_j^{(i)} \in M_{n_j}.$$
In the context of Lie theory, the algebra $\cA$ has irreducible
representation (irrep) of dimensions $n_1, n_2, \ldots $.
Then for every index $j$,   we have a decomposition
$$
\cA\subset (I_{s_j}\otimes M_{n_j})\oplus M_q  \subset M_N,
$$
where $q=  N - s_jn_j$.
If the channel (\ref{channel}) is applied to a quantum state
$\rho = (\tilde \rho \otimes \sigma) \oplus O_q$
according to this decomposition, then we verify
\begin{equation}\label{phi-1}
\Phi(\rho) = (\tilde \rho \otimes \sigma_E) \oplus O_q\qquad
\mbox{with $\sigma_E= \sum_{i=1}^k B_j^{(i)} \sigma  B_j^{(i)\dag}$}
\end{equation}
because of the special form of the error operators in this decomposition.
Thus, the state $\tilde \rho \in M_{s_j}$ encoded as above will not be
affected by the errors (noise) and can be easily recovered.
This gives rise to a noiseless subsystem (NS) \cite{Knill,YBG,Kempe,Kribs}.
The situation is particularly pleasant if $n_j = 1$,
i.e., we use the one-dimensional
irreducible representations of $\cA$, so that
\begin{equation} \label{phi-2}
\Phi(\tilde \rho \oplus O_q) = \tilde \rho \oplus O_q.
\end{equation}
In such a case, we get a decoherence free subspace (DFS) 
\cite{Zan1,Zan2,Zan3,Lidar,Kribs}. Needless to say,
we must identify an irrep $j$ that has the maximal multiplicity $s_j$
for maximal coding rate in both cases. 
Now we consider general qudits, each of which 
belongs to the vector space $\mathbb{C}^d$, instead of qubits.
The irrep can protect
$\log_d {s_j}$ qudits of information from the collective rotation error.
It is the purpose of
the present paper to identify such an irrep that attains the maximal multiplicity
achieving the asymptotic coding rate of one.
Noiseless subsystems and decoherence free subspaces for qudits have been
analyzed by several authors \cite{BRS,GLNPS,Byrd,Bishop1,Bishop2,Bishop3} 
but none of them considered the maximal multiplicity.

In this paper, we study QECC for quantum channel with error operators
on $M_d\otimes \cdots \otimes M_d$ ($n$ copies)
of the form $U^{\otimes n}$ with $U \in {\mathrm{SU}}(d)$.
For this purpose
we analyze the structure of the algebra $\cA(d,n)$
generated by the set
$\cE(d,n)
= \{U^{\otimes n}: U \in {\mathrm{SU}}(d)\}$.
It is known (e.g., see \cite{CPW,FH,Wesslen} and references therein)
that the decomposition has the form
\begin{equation}
\label{decomposition}
\cA(d,n) = \bigoplus_{(p_1, \dots, p_d)}
I_{f(p_1, \dots, p_d)} \otimes  M_{g(p_1, \dots, p_d)}
\end{equation}
with the multiplicity (the Frobenius formula)
$$
f(p_1, \dots, p_d) = \frac{n! \prod_{1\le i < j \le d} (p_i - i - p_j + j)}
{\prod_{\ell=1}^d (p_\ell-\ell+d)!}
$$
and the dimension of the irrep
$$
g(p_1, \dots, p_d) = \prod_{1 \le i < j \le d} \frac{p_i-i-p_j+j}{j-i},
$$
where $p_1 \ge \cdots \ge p_d$ are nonnegative integers arranged in
descending order summing up to $n$.
Each sequence $(p_1, \dots, p_d)$ corresponds to a partition of $n$ into $d$ parts,
equivalently, a Young (tableau) diagram with $n$ boxes.
It is the purpose of this paper
to study which $(p_1, \ldots, p_d)$ maximizes $f(p_1, \ldots,
p_d)$ for a set of $n$ qudits and show it gives the asymptotic encoding 
rate of one.

In Section 2, we summarize the results for the channels on qubits, and
obtain the maximum dimension of the noiseless subsystem
which can be used as the quantum error correction code for the channel.
Then we extend the results to general $d$ in Section 3.
It will be shown
that the code rate, i.e., the number of protected qudits
over the number of physical qudits, always approaches 1 for
a suitable noiseless subsystem.   Moreover, one can determine the maximum
dimension of the noiseless subsystem by solving a non-trivial discrete
optimization problem. In Section 4, the maximum dimension for the case when
$d = 3$ (qutrits) 
is determined by a combination of mathematical analysis and the symbolic
software Mathematica. A conclusion is given in Section 5.
Some technical proofs are collected in Appendix A. 

In addition to the study of quantum error correction,
the results also have implications in Lie theory and
representation theory concerning the algebra
$\cA(d,n)$. In fact, researchers have considered the asymptotic
behavior of $\log_d(f(p_1, \dots, p_d))/n$ and the maximum
value of $f(p_1, \dots, p_d)$; see \cite{Wesslen} and references therein.

\section{Collective rotation on qubits}

It is known \cite{Viola} (see also \cite{Holbrook}) that
$$\cA(2,n) = \bigoplus_{0 \le j \le \lf n/2\rf}
I_{f(n-j,j)}\otimes M_{g(n-j,j)},$$
where $f(n-j,j) = {n\choose j} - {n\choose j-1}$
and $g(n-j,j) = {n+1-2j}$  for $j = 0, \dots, \lf n/2\rf$,
where $\lf x\rf$ denotes the floor of $x \in \IR$, i.e., the
largest integer smaller than or equal to $x$.
Moreover \cite{Zan1,Kempe},
\begin{equation}\label{lim-2}
\lim_{n \rightarrow \infty} \frac{\log_2 f(n-\lf n/2\rf, \lf n/2\rf)}{n} = 1
\end{equation}
so that the code rate, i.e., the number of
protected qubits over the number $n$ of physical qubits, approaches to 1
as $n$ approaches infinity. In case $n$ is even, 
$f(n-n/2,n/2) = n!/(n/2)!(n/2+1)!$ gives the multiplicity of
the 1-dimensional irrep \cite{Zan1,Kempe}.

Despite the limit equation (\ref{lim-2}),
the maximum value of $f(n-r,r)$ is not attained at $r = \lf n/2\rf$.
Actually, we have the following result providing the maximum
dimension of the noiseless subsystem.

\begin{theorem} \label{2.1}
Let $r^*$ be such that
$$f(n-r^*,r^*) = \max_{1 \le r \le n/2} f(n-r,r) =
\max\left\{{n\choose r} - {n\choose r-1}: 1 \le r \le n/2\right\}.$$
Then
$$r^* = \left\lfloor \frac{(n+2) - \sqrt{n+2}}{2} \right\rfloor.$$
\end{theorem}

\noindent
\it Proof. \rm
Note that for $r \le n/2$,
\begin{eqnarray*}
\lefteqn{\left\{{n\choose r} - {n\choose r-1}\right\} -
\left\{{n\choose r-1} - {n\choose r-2}\right\}}\\
&=&  {n\choose r} - 2 {n\choose r-1} + {n\choose r-2} \\
&=& \frac{n!}{r!(n-r+2)!} \{(n+1)(n+2) - 4(n+2)r + 4r^2\},
\end{eqnarray*}
which is nonnegative as long as $r \le \left\{(n+2) - \sqrt{n+2}\right\}/2$.
Thus, the maximum dimension of the noiseless subsystem is attained at
$(n-r^*,r^*)$ with
$$r^* = \left\lfloor \frac{(n+2) - \sqrt{n+2}}{2} \right\rfloor$$
as asserted.
\qed
\\

It is easy to check that if $f(n-r^*,r^*)$ is maximum for a given $n$, then
$$\max\{f(n+1-r^*,r^*), f(n-r^*, r^*+1)\} = \max\{f(p_1,p_2):
p_1 \ge p_2 \ge 1, \ p_1+p_2 = n+1\}.$$

We list the first few values of $r^*$ and $f(n-r^*,r^*)$:
$$
\begin{array}{|r|r|r|r|r|r|r|r|r|r|r|r|r|r|}
\hline
n:   & 3 & 4 & 5 & 6 & 7 & 8 & 9 & 10 & 11 & 12 & 13 & 14 & 15 \\ 
\hline
r^*: & 1 & 1 & 2 & 2 & 3 & 3 & 3 & 4 & 4 & 5 & 5 & 6 & 6\\ 
\hline
f(n-r^*,r^*): & 2 & 3 & 5 & 9 & 14 & 28 & 48 & 90 & 165 & 297 & 572 & 1001 & 2002\\ 
\hline
\lf \log_2 (f(r^*)) \rf: & 1 & 1 & 2 & 3 & 3 & 4 & 5 & 6 & 7 & 8 & 9 & 9 & 10\\ 
\hline
\end{array}
$$

For definiteness, we give an example of NS for 3-qubit encoding \cite{GLNPS}.
For this purpose, we introduce the Young-Yamanouchi basis
\begin{eqnarray}\label{eq:yy}
\begin{aligned}
 \young(12,3) & \left\{ \begin{array}{l}
\begin{aligned}
 &  \frac{1}{\sqrt 6}\left(- [ud + du] u + 2 [uu] d \right) \\
 & \frac{1}{\sqrt 6}\left(2 [dd] u - [ud + du] d \right)
\end{aligned}
\end{array} \right. \\
\young(13,2) & \left\{ \begin{array}{l}
\begin{aligned}
& \frac{1}{\sqrt 2}(ud - du) u \\
& \frac{1}{\sqrt 2}(ud - du) d
\end{aligned}
\end{array} \right. \\
\young(123) & \left\{  \begin{array}{l}
\begin{aligned}
& uuu \\
& \frac{1}{\sqrt 3}(uud + udu + duu) \\
& \frac{1}{\sqrt 3}(ddu + dud + udd) \\
& ddd
\end{aligned}
\end{array} \right.
\end{aligned}
\end{eqnarray}
where $u=|0 \>$ and $d = |1 \>$. The Young diagram in the equation shows
respective symmetry under the symmetry group $S_3$. There are two 2-d
irreps since there are two ancestors, $S=0$ and $S=1$, for $n=2$ and
the Young diagram for $n=3$ shows which ancestor the given 2-d irrep has. 
The encoding unitary matrix $U_E$ is obtained by juxtaposition of
these vectors as
 \begin{eqnarray}
  U_E = \begin{pmatrix}
  \vdots & \vdots & \vdots \\
   \young(12,3) & \young(13,2) & \young(123)\\
   \vdots & \vdots & \vdots \\
  \end{pmatrix}.
\end{eqnarray}
Then a collective rotation operator $U^{\otimes 3}$ is block diagonalized as
\begin{equation}
U_E^{\dagger} U^{\otimes 3} U_E =( I_2 \otimes U_2) \oplus U_4,
\end{equation}
where $U_2 \in {\rm SU}(2)$ and $U_4 \in {\rm SU}(4)$.

In general, it is not easy to implement the encoding and decoding scheme
of QECC of the maximum dimension or of dimension $f(n-\lf n/2\rf,\lf n/2\rf)$ in
the limit equation (\ref{lim-2}). In \cite{QECC3}, a recursive
encoding and
decoding scheme protecting $k$ qubits from collective rotation errors
using $2k+1$ physical qubits 
was demonstrated, leading to an asymptotic code rate of $1/2$. 
Similarly a recursive scheme protecting $k$ qudits using $dk+1$ physical 
qudits with asymptotic code rate of $1/d$ was found in \cite{GLNPS}. 

In fact, the existence of noiseless subsystem/decoherence free subspace
is not a consequence of the unitary group but that of the symmetry group \cite{GLNPS}.
Any collective error symmetric under the permutation of qubit indices can be
shown to have noiseless subsystem by employing the Young-Yamanouchi basis (\ref{eq:yy}). In fact,
let us replace the unitary error operator $U$ by an arbitrary element $M \in \mathrm{GL}(2,\IC)$,
where
$$
M = \left( \begin{array}{cc}
a&b\\
c&d
\end{array} \right).
$$ 
The collective error $M^{\otimes 3}$ for the 3-qubit system is block diagonalized 
by the basis Eq.~(\ref{eq:yy}) as
$$
U_E^{\dagger}  M^{\otimes 3} U_E = (I_2 \otimes M_2) \oplus M_4, 
$$
where 
$$
M_2 = \det M  \times \left(\begin{array}{cc}
d&c\\
b&a
\end{array} 
 \right) \in \mathrm{GL}(2,\IC),
$$
and $M_4 \in \mathrm{GL}(4,\IC)$.
The subsystem corresponding to $I_2$ in
the above expansion is immune to the collective noise and hence noiseless.  
The reader should be referred to \cite{GLNPS} for the case of qudits.

\section{Asymptotic behavior of $f(p_1, \dots, p_d)$}

In this section, we study the asymptotic behavior of $f(p_1, \dots, p_d)$.
For every 
partition $(p_1, \dots, p_d)$ of $n=\sum_{i=1}^d p_i$ with $p_1 \geq p_2 \geq
\ldots p_d >0$, we can get a NS of dimension $f(p_1, \dots, p_d)$.
Here we show that the asymptotic code rate is 1
for a suitable choice of noiseless subsystem, which was known 
to some researchers. Here we give a proof of this fact.

\begin{theorem}\label{thm2}
Suppose $d$ is a positive integer. Then for $p_1 = \dots = p_d = k$, we have
$$\lim_{k\rightarrow \infty} \frac{\log_d f(p_1,\dots,p_d)}{dk} = 1.$$
\end{theorem}

The proof is given in Appendix~\ref{a1}.
Theorem \ref{thm2} states that the choice $p_1 = p_2 = \ldots = p_d = k$ gives the
asymptotic encoding rate of 1. Note, however, that
the maximum value $f(p_1, \dots, p_d)$ for $n=dk$
usually does not occur at
$p_1 = \cdots = p_d = k$.
Moreover, $n$ is not necessarily a multiple of $d$ in general.
In general, it is not easy to determine the maximum
dimension of the noiseless subsystem, equivalently,
to maximize
$$f(p_1, \dots, p_d) = \frac{n! \prod_{1\le i < j \le d} (p_i - i - p_j + j)}
{\prod_{\ell=1}^d (p_\ell-\ell+d)!} .$$
This is a highly non-trivial discrete optimization problem.
Nevertheless, we have the following necessary conditions for the
optimal solution.

\medskip
\it Suppose $f(p_1, \dots, p_d)$ attains the maximum value at $(p_1^*, \dots, p_d^*)$.
Then
$$f(p_1^*+\delta_1, \cdots,
p_m^*+ \delta_d) \le f(p_1^*, \dots, p_d^*)$$
for any $(\delta_1, \dots, \delta_d)$ such that
the vector has only two nonzero entries equal to $1$ and $-1$,
and $p_1^* + \delta_1 \ge \cdots \ge p_d + \delta_d \ge 0$.

\medskip\rm
\noindent
This gives rise to $2{d \choose 2}$ inequalities on $p_1^*, \dots, p_m^*$.
When $d = 2$, the inequalities reduces to a quadratic expression
as shown in the proof of Theorem \ref{2.1}.
When $d = 3$, we will show in the next section that the
$2{d \choose 2} = 6$ inequalities actually determine the optimal solution.
Additional properties of the optimal solutions will be obtained.

We note {\it en passant} that a DFS, corresponding to a square Young diagram
with $d$ rows and $k$ columns ($n=dk$), gives an asymptotic code rate
of 1 even though it does not attain the maximum multiplicity $f$.
Nevertheless, DFS has an advantage of protecting quantum information
from erasure error \cite{MB,MZLWJ,CGL} if there is only one missing qudit. 
Although there are quantum codes that can correct more missing qudits,
these codes require a huge code size, in the form of one-way computing
\cite{vbr}
or surface codes \cite{bs,sb}, to compensate up to $50\%$ qudits loss.

\section{Maximum dimension of a noiseless subsystem for qutrits}

Denote by $\left\lceil x \right\rceil$ the smallest integer $p$ such that
$p \ge x$.
For $d=3$, we have the following result giving the complete information
of the partitions $(p_1^*,p_2^*,p_3^*)$ which yield maximum
$f(p_1,p_2,p_3)$.

\begin{theorem} \label{4.1}
The maximum value $f(p_1,p_2,p_3)$ for $p_1 + p_2 + p_3 = n \ge 3$
can be determined as follows.
\begin{itemize}
\item[{\rm (a)}] If $n = 3k$,
then $f$ has a unique maximum at $(p_1^*,p_2^*,p_3^*) = (k+\left\lceil r_0\right\rceil,k,k-\left\lceil r_0\right\rceil)$ with
$$r_0 = \frac{1}{2} \left(-3+\sqrt{3+3 k+\sqrt{12+20 k+9 k^2}}\, \right).$$
\item[ {\rm (b)}]
If $n = 3k+1$,  then $f$ has a maximum at $(p_1^*,p_2^*,p_3^*) = (k+\left\lceil r_3\right\rceil+1,k,k- \left\lceil r_3\right\rceil)$
with
$$r_3 = \frac14\(-8+\sqrt{40+24k}\) .$$
The maximum is attained at a unique triple $(p_1^*,p_2^*,p_3^*)$
unless $r_3$ is an integer,
equivalently, when 
$k = 1 + 8 q + 6  q^2 $ or $9+16q+6q^2$, for some integer $q\ge 0 $.
In the latter case, $f$ also has a  maximum at $(p_1^*,p_2^*,p_3^*) =(k+\left\lceil r_3 \right\rceil+2,k,k-\left\lceil r_3 \right\rceil -1)$.
\item[{\rm (c)}] If $n = 3k+2$,  let
$$
\begin{array}{ll}
r_1=\displaystyle\frac14\(-10+\sqrt{60+24k}\), &  r_2=\displaystyle\frac14\(-9+\sqrt{49+24k}\), \\[4mm] r_3=\displaystyle\frac14\(-8+\sqrt{40+24k}\), & r_4=\displaystyle\frac14\(-7+\sqrt{49+24k}\).
\end{array}$$
Note that $r_1<r_2<r_3<r_4$.
Then $f$ has a maximum at
$$(p_1^*,p_2^*,p_3^*) =
\left\{\begin{array}{ll}\(k+1+\left\lceil r_3\right\rceil,k+1,k-\left\lceil r_3\right\rceil\)&\mbox{ if } \    \left\lceil r_1\right\rceil \le r_2 \mbox{ or  } \left\lceil r_1\right\rceil \ge r_4 ,\\&\\
\(k+ 2+\left\lceil r_1 \right\rceil ,k,k-\left\lceil r_1\right\rceil \)&\mbox{ if }\
r_2  \le  \left\lceil r_1\right\rceil \le r_4 .
\end{array}\right.$$
Furthermore, $f\(k+1+\left\lceil r_3\right\rceil,k+1,k-\left\lceil r_3\right\rceil\)=f\(k+ 2+\left\lceil r_1 \right\rceil ,k,k-\left\lceil r_1\right\rceil \)$ if and only if $r_2$ or $r_4$ is an integer. We have
\begin{enumerate}
\item $r_2$ is an integer if and only if $k=
  5 + 13 q + 6 q^2$ or $ 10 + 17 q + 6 q^2$ for some integer $q\ge 0$. In these  cases, $\left\lceil r_1\right\rceil =r_2$.
  \item $r_4$ is an integer if and only if $k=
 7 q + 6 q^2$ or $ 3 + 11 q + 6 q^2$ for some integer $q\ge 0$.  In these  cases, $\left\lceil r_1\right\rceil =r_4$.
 \end{enumerate}
\end{itemize}
\end{theorem}

The proof of Theorem \ref{4.1} is given in Appendix~\ref{a2}. 
It is based on the analysis of the following necessary conditions 
on $(p_1, p_2,p_3)$ for $f(p_1, p_2, p_3)$ to be the maximum:
$$\begin{array}{rcll}
f(p_1,p_2,p_3)&-&f(p_1+1,p_2-1,p_3) \geq 0 \qquad&(1)\\[2mm]
f(p_1,p_2,p_3)&-&f(p_1,p_2-1,p_3+1) \geq 0 \qquad&(2)\\[2mm]
f(p_1,p_2,p_3)&-&f(p_1-1,p_2+1,p_3) \geq 0 \qquad&(3)\\[2mm]
f(p_1,p_2,p_3)&-&f(p_1,p_2+1,p_3-1) \geq 0 \qquad&(4)\\[2mm]
f(p_1,p_2,p_3)&-&f(p_1+1,p_2,p_3-1) \geq 0 \qquad&(5)\\[2mm]
f(p_1,p_2,p_3)&-&f(p_1-1,p_2,p_3+1) \geq 0 \qquad&(6)
\end{array}$$
The proof shows that these conditions are actually sufficient.

\medskip
Using Theorem \ref{4.1} and an additional technical lemma, 
one can show that there are close connections
between the partition(s) $(p_1,p_2,p_3)$ attaining maximum $f(p_1,p_2,p_3)$
and those attaining $f(\hat p_1,\hat p_2,\hat p_3)$
with $\hat p_1 + \hat p_2 + \hat p_3 = p_1 + p_2 + p_3 + 1$.
In particular, one can construct the optimal partition as follows.
\begin{itemize}
\item[\rm (a)] For $n = 3k$,
if $f(k+r,k,k-r)$ is the maximum,
then either $f(k+r+1,k,k-r)$ or $f(k+r,k,k-r+1)$
will be the maximum for $n = 3k+1$.
It is possible that
$f(k+r+1,k,k-r)$ and $f(k+r,k,k-r+1)$ are equal
and are both maximum.

\medskip
\item[\rm (b)]
For $n = 3k+1$, if $f(k+r+1,k,k-r)$ is the maximum,
then $f(k+r+2,k,k-r)$, $f(k+r+1,k+1,k-r)$ or $f(k+r+1,k,k-r+1)$
will be the maximum for $n = 3k+2$.

For instance,
$f(2,1,1)$ and $f(3,1,1)$ are both maxima for $n = 4$ and $n = 5$,
respectively;
$f(4,2,1)$ and   $f(4,3,1)$ are maxima for $n = 7$ and $n = 8$,
respectively;
$f(21,16,12)$  and $f(21,16,13)$ are maxima for
$n = 49$ and $n =50$, respectively.

It is possible that
$f(k+r+2,k,k-r) = f(k+r+1,k+1,k-r)$ is maximum
or $f(k+r+1,k+1,k-r) = f(k+r+1,k,k-r+1)$ is maximum,
but $f(k+r+2,k,k-r)$ and $f(k+r+1,k,k-r+1)$ cannot be both maxima.
Furthermore, if there is $\hat r$ such that
$f(k+\hat r+1,k,k-\hat r) = f(k+\hat r+2,k,k-\hat r-1)$
is the maximum when $n = 3k+1$,
then $f(k+\hat r+2,k,k-\hat r)$ is the maximum when $n = 3k+2$.

\medskip
\item[\rm (c)] For $n = 3k+2$, if $f(k+r+1,k+1,k-r)$
or $f(k+r+2,k,k-r)$
is the maximum, then $f(k+r+2,k+1,k-r)$
will be the maximum when $n = 3k+3$.
\end{itemize}

\medskip
The above discussions can be summarized as    the following theorem.

\begin{theorem} \label{mbm}
{\bf (Maximum begets maximum)} Let $f^*(n)$ denote the maximum value of $f(p_1,p_2,p_3)$ for $p_1+p_2+p_3=n$. Then we have
\begin{itemize}
\item[\rm (a)] If $f^*(n) = f(p_1,p_2,p_3)$, then
$$f^*(n+1)=\max\{f(p_1+1,p_2,p_3),\ f(p_1,p_2+1,p_3),\ f(p_1,p_2,p_3+1)\}.$$
\item[\rm (b)] If $f(p_1,p_2,p_3)=f^*(n+1)$, then
$$f^*(n)=\max\{f(p_1-1,p_2,p_3),\ f(p_1,p_2-1,p_3),\ f(p_1,p_2,p_3-1)\}.$$
\item[\rm (c)] If $f^*(n) =f(p^1_1,p^1_2,p^1_3)=f(p^2_1,p^2_2,p^2_3)$ for two distinct $(p^1_1,p^1_2,p^1_3)$ and $(p^2_1,p^2_2,p^2_3)$, then
$$f^*(n+1) =f(p_1,p_2,p_3) \quad\hbox{with}\quad \hbox{$p_i=\max\{p^1_i,p^2_i\}$ for $i=1,2,3$.}$$
\end{itemize}
\end{theorem}

The proof of the theorem is given in Appendix~\ref{a4}.

\section{Conclusion}

We study QECC for quantum channel with error operators
on $M_d\otimes \cdots \otimes M_d$ ($n$ copies)
of the form $U^{\otimes n}$ with $U \in {\rm{SU}}(d)$
by analyzing the structure of algebra $\cA(d,n)$
generated by the set
$\cE(d,n) = \{ U^{\otimes n}: U \in {\rm{SU}}(d)\}$.
We show
that the code rate, i.e., the number of protected qudits
over the number of the physical qudits, always approaches 1  for
a suitable noiseless subsystem.  Moreover, we have determined the maximum
dimension of the noiseless subsystem by solving a non-trivial
discrete optimization  problem. The maximum dimension for the cases
when $d = 2, 3$ are determined by
a combination of mathematical analysis and the symbolic software
Mathematica. Additional properties  of the partition
$(p_1, \dots, p_d)$ of $n$ attaining the maximum dimension
$f(p_1, \dots, p_d)$ are obtained when $d = 3$.

In general, it is not easy to construct a noiseless subsystem with
the maximum dimension $f(p_1^*, \dots, p_d^*)$ for a large $n$
 as shown in the cases when $d = 2$. For a large $n$, the encoding
 and decoding unitary matrices are elements of $U(2^n)$ and it is 
impossible to write down these exponentially large unitary matrices. 
It seems to us that recursive
implementation making use of encoding and decoding circuits for
a small $n$ to build up those for a larger $n$ is the only sensible way
for physical implementation. In fact, we found such recursive schemes for
qubits \cite{QECC3} and qudits \cite{GLNPS}. However, in these cases
we have to sacrifice the code rate, which is
$1/2$ for qubits and $1/d$ for qudits in the recursive schemes. 
Implementation of encoding and decoding circuits for the maximal
encoding scheme remains a challenging future research.

\section*{Acknowledgment}
The research of Li was supported dy a USA NSF grant, a HK RGC grant.
He is an affiliate member of the Institute for Quantum Computing,
University of Waterloo; an honorary professor of the Shanghai University
and the University of Hong Kong.
The research of Nakahara was supported by Grants-in-Aid for Scientific
Research from the Japan Society for the Promotion of Science (Grant
Nos. 23540470, 24320008 and 26400422). He is grateful to Qing-Wen Wang
and Xi Chen for warm hospitality extended to him while he was staying
at Shanghai University, where a part of this work was done.
The research of Poon was supported by a USA NSF grant and a HK RGC grant.
The research of Sze was supported by a HK RGC grant PolyU 502512.
The authors want to thank Utkan G\"{u}ng\"{o}rd\"{u} for some helpful 
discussion
concerning the decomposition in (\ref{decomposition}). We would like to thank
Paolo Zanardi for drawing our attention to Refs. \cite{Zan1,Zan2,Zan3} and
useful comments that improved our manuscript.

\appendix

\section{Proofs of Theorems}

\subsection{Proof of Theorem~\ref{thm2}}\label{a1}

By the Frobenius formula \cite{FH,CPW,Wesslen},
$$f(p_1,\dots,p_d) = \frac{(dk)! \prod_{\ell=1}^{d-1} \ell!} {\prod_{\ell=1}^d (k+d-\ell)!}
= f_1(k)f_2(k),$$
where
$$f_1(k) = \frac{(dk)!}{(k!)^d} \qquad \hbox{ and } \qquad
f_2(k) =  \frac{ \prod_{\ell=1}^{d-1} \ell! }{\prod_{\ell=1}^d ((k+1)\cdots (k+d-\ell))}.$$
Now, by Stirling's formula, we obtain
\begin{eqnarray*}
\lim_{k\rightarrow \infty} \frac{\log_d f_1(k)}{dk}
&=& \lim_{k\rightarrow \infty} \frac{\ln(dk)! - d \ln k!} {(\ln d)dk}\\
&=& \lim_{k\rightarrow \infty} \frac{ dk(\ln d) + O(\ln(dk)) - O(\ln k )} {(\ln d) dk} = 1,
\end{eqnarray*}
and
$$\lim_{k\rightarrow \infty} \frac{\log_d(f_2(k))}{dk} =
\lim_{k\rightarrow \infty} \frac{\log_d( \prod_{\ell=1}^{d-1} \ell!) -
\sum_{i=1}^d \sum_{j=1}^{d-i} \log_d(k+j)} {dk} = 0.$$
The result follows. \qed

\subsection{Proof of Theorem~\ref{4.1}}\label{a2}

Consider the following terms:
$$\begin{array}{rcll}
f(p_1,p_2,p_3)&-&f(p_1+1,p_2-1,p_3)\qquad&(1)\\[2mm]
f(p_1,p_2,p_3)&-&f(p_1,p_2-1,p_3+1)\qquad&(2)\\[2mm]
f(p_1,p_2,p_3)&-&f(p_1-1,p_2+1,p_3)\qquad&(3)\\[2mm]
f(p_1,p_2,p_3)&-&f(p_1,p_2+1,p_3-1)\qquad&(4)\\[2mm]
f(p_1,p_2,p_3)&-&f(p_1+1,p_2,p_3-1)\qquad&(5)\\[2mm]
f(p_1,p_2,p_3)&-&f(p_1-1,p_2,p_3+1)\qquad&(6)
\end{array}$$
Clearly, if $f(p_1,p_2,p_3)$ attains the maximum, then each of the
above terms is nonnegative.

\medskip
\noindent
(a) When $n = 3k$, there are three cases (a.1) $p_2 > k$, (a.2) $p_2<k$,
and (a.3) $p_2 = k$.
We use Mathematica \cite{Math} to show the following.
\begin{itemize}
\item[\rm (a.1)] If $(p_1,p_2,p_3) = (k+a+ b,k+a,k-2a-b)$ with $a > 0$ and $b \ge 0$,
the sum of  (1) and (2) is negative and
so $f$ cannot attain the maximum in this case.

\item[\rm (a.2)] If $(p_1,p_2,p_3) = (k+2a+b,k-a,k-a-b)$ with $a > 0$ and $b \ge 0$,
the either (3) or (4) is negative and
so $f$ cannot attain the maximum in this case.

\item[\rm (a.3)] Let $(p_1,p_2,p_3) = (k+r,k,k-r)$ and define $ r_0 \equiv
\frac{1}{2} \left(-3+\sqrt{3+3 k+\sqrt{12+20 k+9 k^2}}\right)$.
Then $(5)$ and $(6)$ are both nonnegative if and only if
$\left\lceil r_0 \right\rceil \leq r \leq \left\lceil r_0\right\rceil+1$.
\end{itemize}

Moreover, $r_0$ in (a.3) is not an integer for all $k\ge 1$.
Thus, $f(k+r,k,k-r)$ will attain its maximum when $r = \left\lceil r_0 \right\rceil$.
Therefore, $f$ has a unique maximum
$f(k+r,k,k-r)$ with $r = \left\lceil r_0 \right\rceil$ when $n = 3k$ as stated.

\medskip\noindent
(b) When $n = 3k+1$, we also have three cases (b.1) $p_2>k$,
(b.2) $p_2<k$ and (b.3) $p_2=k$. We use Mathematica \cite{Math} to show the following.
\begin{itemize}
\item[\rm (b.1)] If $(p_1,p_2,p_3) = (k+1+a+b,k+a,k-2a-b)$ with $a > 0$ and $b \ge 0$,
the sum of (1) and (2) is negative and so $f$ cannot attain the maximum in this case.

\item[\rm (b.2)] If $(p_1,p_2,p_3) = (k+1+2a+b,k-a,k-a-b)$ with $a > 0$ and $b \ge 0$,
then either (3) or (4) is negative and so $f$ cannot attain the maximum in this case.

\item[\rm (b.3)] Let $(p_1,p_2,p_3) = (k+1+r,k,k-r)$ and define $r_3 \equiv
\frac{1}{4}(-8 + \sqrt{40+24k})$. Then (5) and (6) are both
nonnegative if and only if $\lceil r_3 \rceil \leq r \leq \lceil r_3 \rceil+1$.
\end{itemize}
There is a subtlety that does not exist for the case (a).
We show that $r_3$ is a positive integer if and only if
$k = 1 + 8 q + 6  q^2 $ or $9+16q+6q^2$, for some integer $q\ge 0$.
In this case, The term $(5)$ is equal to zero when $(p_1,p_2,p_3) = (k+1+r,k,k-r)$.
Therefore, $f(k+1+r,k,k-r)=f(k+2+r,k,k-r-1)$ and $f$ has a maximum value at $(p_1,p_2,p_3)=(k+1+r,k,k-r)$ and $(k+2+r,k,k-r-1)$.

\medskip\noindent
(c) When $n = 3k+2$, we use Mathematica \cite{Math} to show the following.
\begin{itemize}
\item[\rm (c.1)] If $(p_1,p_2,p_3) = (k+a+b,k+a,k+2-2a-b)$ with $a > 1$ and $b \ge 0$,
the sum of (1) and (2) is negative and
so $f$ cannot attain the maximum in this case.


\item[\rm (c.2)] If $(p_1,p_2,p_3) = (k+2+a+b,k-a,k-a-b)$ with $a \ge 1$ and $b > 0$,
the sum of  (3) and (4) is negative and
$f$ cannot attain the maximum in this case.
\end{itemize}
The cases (c.1) and (c.2) show that the maximum of $f(p_1,p_2,p_3)$ can occur
only at
$p_2=k$ or $k+1$. We show by Mathematica \cite{Math} that
\begin{itemize}
\item[\rm (c.3)] For $r \ge 0$,
$f(k+2+r,k,k-r)\le f(k+2+\left\lceil r_1 \right\rceil,k,k-\left\lceil r_1 \right\rceil)$ and
$f(k+1+r,k+1,k-r)\le f(k+1+\left\lceil r_3 \right\rceil,k+1,k-\left\lceil r_3 \right\rceil)$.

\item[\rm (c.4)] For $r\ge 0$ and  $(p_1,p_2,p_3)=(k+2+r,k,k-r)$,
(3) is non-positive if and only if $r\ge r_4 $ and
(4) is non-positive if and only if $ r\le r_2 $.
\end{itemize}
By (c.3), the maximum of $f$ occurs at either $(p_1,p_2,p_3) =
(k+2+\left\lceil r_1 \right\rceil,k,k-\left\lceil r_1 \right\rceil)$
or $(k+1+\left\lceil r_3 \right\rceil,k+1,k-\left\lceil r_3 \right\rceil)$.
Since $0< r_3-r_1=\frac14\(2-\sqrt{60+24k}+\sqrt{40+24k}\)<\dfrac{1}{2}$, we have $\left\lceil r_1 \right\rceil\le \left\lceil r_3 \right\rceil\le \left\lceil r_1 \right\rceil+1$.
Furthermore, by (c.3) and (c.4),
\begin{equation}\label{r1}
\begin{array}{c}
f(k+2+\left\lceil r_1 \right\rceil,k,k-\left\lceil r_1
\right\rceil)\le f(k+1+\left\lceil r_3 \right\rceil,k+1,k-\left\lceil r_3
\right\rceil)\quad\\
\mbox{  if }\left\lceil r_1 \right\rceil\le r_2\ \hbox{or}\ \left\lceil
r_1 \right\rceil\ge r_4
\end{array}
\end{equation}

Conversely, suppose $r_2\le \left\lceil r_1 \right\rceil\le r_4$. Then both (3) and (4) are nonnegative. Hence,
\begin{eqnarray*}
f(k+2+\left\lceil r_1 \right\rceil,k,k-\left\lceil r_1 \right\rceil)
&\ge& f(k+1+\left\lceil r_1 \right\rceil,k+1,k-\left\lceil r_1 \right\rceil)
\quad \hbox{and} \\
f(k+2+\left\lceil r_1 \right\rceil,k,k-\left\lceil r_1 \right\rceil)
&\ge &
f(k+2+\left\lceil r_1 \right\rceil,k+1,k-\left\lceil r_1
\right\rceil-1).
\end{eqnarray*}
Since $\left\lceil r_1 \right\rceil\le \left\lceil r_3 \right\rceil\le \left\lceil
r_1 \right\rceil+1$, it follows that
\begin{eqnarray}\label{r2}
\begin{array}{c}
f(k+2+\left\lceil r_1 \right\rceil,k,k-\left\lceil r_1 \right\rceil)\ge
f(k+1+\left\lceil r_3 \right\rceil,k+1,k-\left\lceil r_3 \right\rceil)\\
\quad\hbox{if}\quad
r_2\le \left\lceil r_1 \right\rceil\le r_4.
\end{array}
\end{eqnarray}
By (\ref{r1}) and (\ref{r2}), $f$ has a unique maximum if neither $r_2$ nor $r_4$ is an integer.

Suppose $r_2$ is an integer for some $k$. Then we have
$$
k=\left\{ \begin{array}{ll}
\frac{4}{3} + 9 q + 6 q^2&\mbox{ if }r_2=3q ,\\[2mm]
5 + 13 q + 6 q^2&\mbox{ if }r_2=3q+1 , \\[2mm]
10 + 17 q + 6 q^2&\mbox{ if }r_2=3q+2 .
\end{array}
\right.
$$
Thus, $r_2$ is an integer for some integer $k$ if and only if $k=5 + 13 q +
6 q^2$ or $10 + 17 q + 6 q^2$ for some integer $q\ge 0$.
For $k=5 + 13 q + 6 q^2$,
$r_1=\frac14 \(-10 + \sqrt{180 + 312 q + 144 q^2}\)$. We have
\begin{eqnarray*}
3+12q = -10 + \sqrt{(13+12q)^2}
&\le& -10 + \sqrt{180+312 q + 144 q^2}\\
&\le& -10 + \sqrt{(14+12q)^2}
= 4+12q.
\end{eqnarray*}
So
$$\frac{3}{4} + 3q = \frac{1}{4} (3+12q) \le r_1 \le \frac{1}{4}(4+12q) = 1 + 3q
\quad\Longrightarrow\quad
\left\lceil r_1 \right\rceil=3q+1=r_2.$$
For $k=10 + 17 q + 6 q^2$,
$r_1=\frac14 \(-10 + \sqrt{300 + 408 q + 144 q^2}\)$. We have
\begin{eqnarray*}
7+12q = -10 + \sqrt{(17+12q)^2}
&\le& -10 + \sqrt{300 + 408 q + 144 q^2}\\
&\le &-10 + \sqrt{(18+3q)^2}
= 8 + 12q.
\end{eqnarray*}
So
$$\frac{7}{4} + 3q = \frac{1}{4}(7+12q) \le r_1 \le \frac{1}{4}(8+12q)
= 2 + 3q
\quad\Longrightarrow\quad
\left\lceil r_1 \right\rceil=3q+2=r_2.$$
The proof for the case when $r_4$ is an integer is similar.
\qed


\subsection{Proof of Theorem \ref{mbm}}\label{a4}

We first prove the following Lemma.

\begin{lemma}\label{lem1}
Let $r_0, r_1,r_2,r_3$, and $r_4$ be the numbers defined in Theorem \ref{4.1}
and let
\begin{eqnarray*}
\hat r_0
&=&\frac{1}{2} \left(-3+\sqrt{3+3 (k+1)+\sqrt{12 + 20 (k+1) + 9 (k+1)^2}}\right) \\
&=&\frac{1}{2} \left(-3+\sqrt{6+3 k+\sqrt{41 + 38 k + 9 k^2}}\right).
\end{eqnarray*}
Then
 \begin{equation}\label{rs}
r_1< r_3< r_0< \hat r_0
< r_1 +1 < r_3 + 1.
 \end{equation}
Hence, $\{\left\lceil r_0 \right\rceil,\left\lceil \hat r_0 \right\rceil\}
\subseteq \{\left\lceil r_3\right\rceil, \left\lceil r_3\right\rceil + 1\}$.
Furthermore,
$\left\lceil r_0\right\rceil = \left\lceil r_3\right\rceil + 1$
if $r_3$ is an integer
and $\lceil \hat r_0 \rceil =  \left\lceil r_1\right\rceil + 1$
if $r_2 \le \lceil r_1 \rceil \le r_4$.
\end{lemma}

\noindent
\it Proof. \rm
With the help of Mathematica \cite{Math}, we can show that
\begin{eqnarray*}
\lefteqn{\(r_0-r_3\)
\left(1+\sqrt{10+6 k}+\sqrt{3+3 k+\sqrt{12+k (20+9 k)}}\right)}
\\
& &\times
\left(6+3 k+\sqrt{12+k (20+9 k)}+2 \sqrt{3+3 k+\sqrt{12+k (20+9 k)}}\right)\\
& &\times
\left(3+k+\sqrt{12+k (20+9 k)}\right.\\
& & \left. +\sqrt{12+k (20+9 k)} \sqrt{3+3 k+\sqrt{12+k (20+9 k)}}\right)\\
&=& 2 \left[3 \left(13+4 \sqrt{12+k (20+9 k)}+8 \sqrt{3+3 k+\sqrt{12+k (20+9 k)}}\right)\right.\\
& & +k \left\{10 \left(11+2 \sqrt{12+k (20+9 k)}+4 \sqrt{3+3
k
+\sqrt{12+k (20+9 k)}}\right)\right.\\
& &+\left.\left.k \left(95+27 k+9 \sqrt{12+k (20+9 k)}+18 \sqrt{3+3 k+\sqrt{12+k (20+9 k)}}\right)\right\}\right] \\
&>&0.
\end{eqnarray*}
\begin{eqnarray*}
\lefteqn{(\hat r_0- r_0)\left(\sqrt{3+3 k+\sqrt{12+k (20+9 k)}}+\sqrt{6+3 k+\sqrt{41+k (38+9 k)}}\right)} \\
& & \times \left(3+\sqrt{12+k (20+9 k)}+\sqrt{41+k (38+9 k)}\right) \\
  &=&19+9 k+3 \sqrt{41+k (38+9 k)}>0,
\end{eqnarray*}
and
\begin{eqnarray*}
\lefteqn{\(r_1 +1- \hat r_0 \)\left(\sqrt{15+6 k}+\sqrt{6+3 k+\sqrt{41+k (38+9 k)}}\right) }\\
& & \times \left(9+3 k+\sqrt{41+k (38+9 k)}\right) \\
 &=&20+8 k >0
\end{eqnarray*}
Hence, Eq. (\ref{rs}) follows.
\qed

Now we are ready to prove the theorem.
Notice that by Theorem \ref{4.1},
$$
f^*(n) = \left\{ \begin{array}{ll}            
f(k+\lceil r_0\rceil,k,k-\lceil r_0\rceil) & \hbox{if } n = 3k, \\[1mm]
f(k+ 1 +\lceil r_3\rceil,k,k-\lceil r_3\rceil) & \hbox{if } n = 3k+1, \\[1mm]
f(k+1+\lceil r_3\rceil,k+1,k-\lceil r_3\rceil)
& \hbox{if } n = 3k+2 \hbox{ with }
\left\lceil r_1\right\rceil \le r_2\ \mbox{  or } \left\lceil r_1\right\rceil \ge r_4, \\[1mm]
f(k+2+\lceil r_1\rceil,k,k-\lceil r_1\rceil)
& \hbox{if } n = 3k+2 \hbox{ with }
r_2 \le \left\lceil r_1\right\rceil \le r_4,\\[1mm]
f(k+\lceil \hat r_0\rceil,k,k-\lceil \hat r_0\rceil) & \hbox{if } n = 3k+3. \\[1mm]
\end{array} \right. 
$$
Furthermore, $f(k+ 1 +\lceil r_3\rceil,k,k-\lceil r_3\rceil) = f(k+ 2 +\lceil r_3\rceil,k,k-1-\lceil r_3\rceil)$ if $r_3$ is an integer.

From $n = 3k$ to $n+1 = 3k+1$,
suppose
$(k+\lceil r_0\rceil,k,k-\lceil r_0\rceil) = (p_1,p_2,p_3)$.
By Lemma \ref{lem1}, $\left\lceil r_0 \right\rceil = \left\lceil r_3\right\rceil$ or $\left\lceil r_3\right\rceil + 1$. Then
$$
(k+1+\lceil r_3\rceil,k,k-\lceil r_3\rceil) = \left\{ \begin{array}{ll}
(p_1+1,p_2,p_3) &  \hbox{if } \lceil r_0 \rceil = \lceil r_3 \rceil, \\[2mm]
(p_1,p_2,p_3+1) &  \hbox{if } \lceil r_0 \rceil = \lceil r_3 \rceil+1.
\end{array} \right. 
$$

\medskip
From $n = 3k+1$ to $n+1 = 3k+2$, suppose
$(k+1+\lceil r_3\rceil,k,k-\lceil r_3\rceil)  = (p_1,p_2,p_3)$.
By Lemma \ref{lem1},
$\left\lceil r_3 \right\rceil = \left\lceil r_1\right\rceil$
or $\left\lceil r_1\right\rceil + 1$. Then
\begin{eqnarray*}
(k+1+\lceil r_3\rceil,k+1,k-\lceil r_3\rceil) &=& (p_1,p_2+1,p_3) \quad \hbox{and} \\[1mm]
(k+2+\lceil r_1\rceil,k,k-\lceil r_1\rceil)  &=& \left\{ \begin{array}{ll}
(p_1+1,p_2,p_3) &  \hbox{if } \lceil r_3 \rceil = \lceil r_1 \rceil, \\[2mm]
(p_1,p_2,p_3+1) &  \hbox{if } \lceil r_3 \rceil = \lceil r_1 \rceil+1.
\end{array} \right. 
\end{eqnarray*}
Further, suppose $r_3$ is an integer, then $\lceil r_3 \rceil = \lceil r_1 \rceil$
and hence we must have $r_2 \le \lceil r_1 \rceil \le r_4$
and $f^*(3k+2) = f(k+2+\lceil r_1\rceil,k,k-\lceil r_1\rceil)$.
If $(p_1,p_2,p_3) = (k+2+\lceil r_3\rceil,k,k-1-\lceil r_3\rceil)$, then
$$(k+2+\lceil r_1\rceil,k,k-\lceil r_1\rceil) = (p_1,p_2,p_3+1).$$
Furthermore, if
$f^*(3k+1) = f(p_1^1,p_2^1,p_3^1) = f(p_1^2,p_2^2,p_3^2)$
with
$(p_1^1,p_2^1,p_3^1) = (k+1+\lceil r_3\rceil,k,k-\lceil r_3\rceil)$
and
$(p_1^2,p_2^2,p_3^2) = (k+2+\lceil r_3\rceil,k,k-1-\lceil r_3\rceil)$,
then
$f^*(3k+2) = f(p_1^2,p_2^1,p_3^1)$.

\medskip
Now from $n = 3k+2$ to $n+1 = 3k+3$,
suppose $\left\lceil r_1\right\rceil \le r_2$
or $\left\lceil r_1\right\rceil \ge r_4$ and
$(k+1+\lceil r_3\rceil,k+1,k-\lceil r_3\rceil) = (p_1,p_2,p_3)$.
By Lemma \ref{lem1}, $\left\lceil \hat r_0 \right\rceil = \left\lceil r_3\right\rceil$ or $\left\lceil r_3\right\rceil + 1$. Then
$$(k+1 + \lceil \hat r_0 \rceil,k+1,k+1-\lceil \hat r_0\rceil)
= \left\{ \begin{array}{ll} 
(p_1,p_2,p_3+1) & \hbox{if } \lceil \hat r_0 \rceil = \lceil r_3\rceil, \\[2mm]
(p_1+1,p_2,p_3) & \hbox{if } \lceil \hat r_0 \rceil = \lceil r_3 \rceil + 1.
\end{array}\right. 
$$
Now suppose $r_2 \le \lceil r_1 \rceil  \le r_4$
and $(k+2+\lceil r_1\rceil,k,k-\lceil r_1\rceil) = (p_1,p_2,p_3)$.
By Lemma \ref{lem1}, $\lceil \hat r_0\rceil = \lceil r_1\rceil + 1$. Then
$$(k+1 + \lceil \hat r_0 \rceil,k+1,k+1-\lceil \hat r_0\rceil) = (p_1,p_2+1,p_3).$$
Furthermore, if
$f^*(3k+2) = f(p_1^1,p_2^1,p_3^1) = f(p_1^2,p_2^2,p_3^2)$
with
$(p_1^1,p_2^1,p_3^1) = (k+1+\lceil r_3\rceil,k+1,k-\lceil r_3\rceil)$
and
$(p_1^2,p_2^2,p_3^2) = (k+2+\lceil r_1\rceil,k,k-\lceil r_1\rceil)$,
then $f^*(3k+3) = f(p_1^2,p_2^1,p_3^1)$.
Thus, the result follows.
\qed

\end{document}